# Size Effects of Ferroelectric and Magnetoelectric Properties of Semi-ellipsoidal Bismuth Ferrite Nanoparticles

Victoria V. Khist[1], Eugene A. Eliseev[2], Maya D. Glinchuk[2], Maxim V. Silibin[3], Dmitry V. Karpinsky[4,1], and Anna N. Morozovska[5,2]

[1] Institute of Magnetism, National Academy of Sciences of Ukraine and Ministry of Education and Science of Ukraine,

[2] Institute for Problems of Materials Science, National Academy of Sciences of Ukraine,
3, Krjijanovskogo, 03142 Kyiv, Ukraine,

[3] National Research University of Electronic Technology "MIET",
Moscow, Zelenograd, Russia

[4] Scientific-Practical Materials Research Centre of NAS of Belarus, Minsk, Belarus

[5] Institute of Physics, National Academy of Sciences of Ukraine,
46, pr. Nauky, 03028 Kyiv, Ukraine

Bismuth ferrite ($BiFeO_3$) is one of the most promising multiferroics with a sufficiently high ferroelectric (FE) and antiferromagnetic transition temperatures, and magnetoelectric (ME) coupling coefficient at room temperature, and thus it is highly sensitive to the impact of cross-influence of applied electric and magnetic fields. According to the urgent demands of nanotechnology miniaturization for ultra-high density data storage in advanced nonvolatile memory cells, it is very important to reduce the sizes of multiferroic nanoparticles in the self-assembled arrays without serious deterioration of their properties. We study size effects of the phase diagrams, FE and ME properties of semi-ellipsoidal $BiFeO_3$ nanoparticles clamped to a rigid conductive substrate. The spatial distribution of the spontaneous polarization vector inside the nanoparticles, phase diagrams and paramagnetoelectric (PME) coefficient were calculated in the framework of modified Landau-Ginzburg-Devonshire (LGD) approach. Analytical expressions were derived for the dependences of the FE transition temperature, average polarization, linear dielectric susceptibility and PME coefficient on the particle sizes for a general case of a semi-ellipsoidal nanoparticles with three different semi-axes *a, b* and height *c*. The analyses of the obtained results leads to the conclusion that the size effect of the phase diagrams, spontaneous polarization and PME coefficient is rather sensitive to the particle sizes aspect ratio in the polarization direction, and less sensitive to the absolute values of the sizes per se.

**Keywords:** multiferroic; structural antiferrodistortion; antiferromagnetic order; nanoparticle; semi-ellipsoidal; nano-island

---

[1] corresponding author, e-mail: dmitry.karpinsky@gmail.com
[2] corresponding author, e-mail: anna.n.morozovska@gmail.com

# 1. INTRODUCTION

## 1.1. Multiferroic BiFeO$_3$ for fundamental studies and advanced applications

Multiferroics, which are ferroics with two or more long-range order parameters, are ideal systems for fundamental studies of the couplings among the ferroelectric polarization, structural antiferrodistortion, and antiferromagnetic order parameters [1, 2, 3, 4, 5]. These couplings are in response of unique physical properties of multiferroics [6]. For instance, biquadratic and linear magnetoelectric (**ME**) couplings lead to intriguing effects, such as a giant magnetoelectric tunability of multiferroics [7]. Biquadratic coupling of the structural and polar and dielectric order parameters, introduced in Refs. [8, 9, 10], are responsible for the unusual behavior of the dielectric, polar and other physical properties in ferroelastics – quantum paraelectrics. The linear-quadratic paramagnetoelectric (**PME**) effect should exist in the paramagnetic phase of ferroics, below the temperature of the paraelectric-to-ferroelectric phase transition, where the electric polarization is non-zero. This effect was observed in NiSO$_4$·6H$_2$O [11], Mn-doped SrTiO$_3$ [12], Pb(Fe$_{1/2}$Nb$_{1/2}$)O$_3$ [13, 14, 15], and Pb(Fe$_{1/2}$Nb$_{1/2}$)O$_3$- PbTiO$_3$ solid solution [16]. Note that PME effect can be expected in many nanosized ferroics, which becomes paramagnetic due to the size-induced transition from the ferromagnetic or antiferromagnetic phase.

BiFeO$_3$ is the one of the most interesting multiferroics with a strong ferroelectric polarization, antiferromagnetism at room temperature as well as enhanced electrotransport at domain walls [17, 18, 19, 20, 21, 22]. Bulk BiFeO$_3$ exhibits antiferrodistortive (AFD) order at temperatures below 1200K; it is ferroelectric (FE) with a large spontaneous polarization below 1100 K and is antiferromagnetic (AFM) below Neel temperature $T_N \approx 650$ K [23, 24]. The pronounced multiferroic properties maintain in BiFeO$_3$ thin films and heterostructures [25, 26, 27, 28, 29]. Despite extensive experimental and theoretical studies of the physical properties of bulk BiFeO$_3$ and its thin films [21 - 23, 30, 31, 32, 33, 34, 35, 36, 37], many important issues concerning the emergence and theoretical background of multiferroic polar, magnetic and various electrophysical properties of BiFeO$_3$ nanoparticles remain almost unexplored [38, 39].

## 1.2. Multiferroelectric nanoparticles. State-of-art

However according to the urgent demands of nanotechnology miniaturization for ultra-high density of data storage in advanced nonvolatile memory cells, it is very important to reduce the sizes of multiferroic nanoparticles in the self-assembled arrays without serious deterioration of their polar, magnetic and ME properties. There are many intriguing and encouraging examples of the polar and dielectric properties conservation, enhancement and modification in ferroelectric nanoparticles. In particular Yadlovker and Berger [40, 41, 42] present the unexpected experimental results, which reveal the enhancement of polar properties of cylindrical nanoparticles of Rochelle salt. Frey and Payne [43], Zhao et al [44] and Erdem et al [45] demonstrate the possibility to control the temperature of the ferroelectric phase transition, the magnitude and position of the dielectric constant maximum for BaTiO$_3$ and PbTiO$_3$ nanopowders and nanoceramics. The studies of KTa$_{1-x}$Nb$_x$O$_3$ nanopowders [46] and nanograin ceramics [47, 48, 49] discover the appearance of new polar phases, the shift of phase transition temperature in comparison with bulk crystals. Strong size effects in SrBi$_2$Ta$_2$O$_9$ nanoparticles have been revealed by *in situ* Raman scattering by Yu et al [50] and by thermal analysis and Raman spectroscopy by Ke et al [51]. The list of experimental studies should be continued, making any new experimental-and-theoretical study of ferroelectric nanoparticles important for both fundamental science and advanced applications.

In particular, the surface and finite size effects impact on the phase diagrams, polar and electrophysical properties of BiFeO$_3$ nanoparticles are very poorly studied [38, 39]. Such study may be very useful for science and advanced applications, because the theory of finite size effects in nanoparticles allows one to establish the physical origin of the polar and other physical properties anomalies, transition temperature and phase diagrams changes appeared with the nanoparticles sizes decrease. In particular, using the continual phenomenological approach Niepce [52], Huang et al [53, 54], Ma [55], Eliseev et al [56] and Morozovska et al [57, 58, 59, 60, 61] have shown, that the changes of the transition temperatures, the enhancement or weakening of polar properties in spherical and cylindrical nanoparticles are conditioned by the various physical mechanisms, such as correlation effect, depolarization field, flexoelectricity, electrostriction, surface tension and Vegard-type chemical pressure.

### 1.3. Research motivation

Nanoparticles of (semi)ellipsoidal shape can be considered as the model objects to study size effects on physical properties of ferroic nano-islands. BiFeO$_3$ nano-islands and their self-assembled arrays can be formed on anisotropic substrates by different low-damage fabrication methods [62, 63, 64]. The particles typically have different in-plane axes due to the anisotropic thermal conductivity of substrate. Recent advances in the production technology of ferroelectrics have resulted in a cost-effective synthesis of these nanoparticles, which are beginning to be used in fabrication of microactuators, microwave phase shifters, infrared sensors, transistor applications, energy harvesting devices etc. A correlation mechanism between the scaling factor, geometry of the nanoparticles and their physical parameters, and related phenomena viz. spontaneous polarization, antiferromagnetic and antiferrodistortive order, width of the domain walls and the domains stability is needed to be further investigated using both experimental methods and theoretical modeling. Most intriguing fundamental issues to be addressed include an estimation of the intrinsic limit for polarization stability, mechanism of domain wall motion, and polarization switching in nanoscale volumes [1-6].

The analyses of the above state-of-art motivated us to study theoretically the size effects influence on FE, AFM and ME properties of semi-ellipsoidal BiFeO$_3$ nanoparticles in the framework of the Landau-Ginzburg-Devonshire (LGD) approach, classical electrostatics and elasticity theory.

## 2. THEORY

### 2.1. Problem statement

It is known that ferroelectricity is a cooperative phenomenon associated with the dipole moments aligned on both short- and long-scale level. This alignment is characterized by a certain transition temperature justified by the temperature-dependent forces which relate to the size effects, dimension of the material, its structural homogeneity etc. It is considered that the size effects are associated either with intrinsic (mainly related to the atomic polarization) or extrinsic (stresses, microstructure, polarization screening etc.) factors [57-60].

Let us consider ferroelectric nanoparticles in the form of semi-elliptical islands precipitated on the rigid conducting substrate electrode. The ellipsoid has different values of semi-axis length, *a*, *b* and *c* along X-, Y- and Z-axis respectively. We denote $\varepsilon_b$ and $\varepsilon_e$ as the dielectric permittivity of ferroelectric background and external media respectively. The one-component ferroelectric polarization $\mathbf{P}(\mathbf{r})$, directed along the crystallographic axis 3 inside the particle, that is parallel to the interface z=0 [**Fig. 1**].

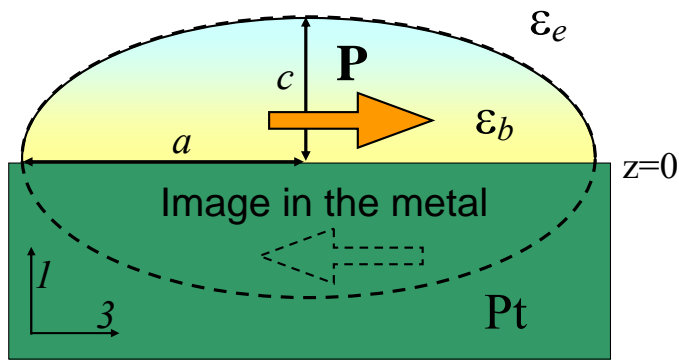

**FIG. 1.** A semi-ellipsoidal uniformly polarized ferroelectric nanoparticle is clamped to a rigid conducting substrate electrode (e.g., Pt). The one-component ferroelectric polarization $\mathbf{P}(\mathbf{r})$ is directed along x-axes. Semi-ellipsoid height is *c* and lateral semi-axes are *a* and *b*.

We can assume that in the crystallographic frame {1, 2, 3} the dependence of the in-plane components of electric polarization on the inner field electric $\mathbf{E}^i$ is linear $P_1 = \varepsilon_0(\varepsilon_b^i - 1)E_1^i$ and $P_2 = \varepsilon_0(\varepsilon_b^i - 1)E_2^i$, where an isotropic background permittivity is relatively small, $\varepsilon_b^i \leq 10$ [65], $\varepsilon_0$ is a universal dielectric constant. Polarization component 3 contains background and soft mode contributions, $P_3(\mathbf{r}, E_3) = P(\mathbf{r}, E_3) + \varepsilon_0(\varepsilon_b^i - 1)E_3^i$. Electric displacement vector has the form $\mathbf{D}^i = \varepsilon_0 \varepsilon_b^i \mathbf{E}^i + \mathbf{P}$ inside the particle and $\mathbf{D}^e = \varepsilon_0 \varepsilon^e \mathbf{E}^e$ outside it; $\varepsilon^e$ is the relative dielectric permittivity of external media. Hereinafter the subscript "*i*" corresponds to the electric field or potential inside the particle, "*e*" – outside the particle.

Inhomogeneous spatial distribution of the ferroelectric polarization component $P_3(\mathbf{r}, E_3)$ can be determined from the Landau-Ginzburg-Devonshire (LGD) type equations inside a nanoparticle,

$$\alpha_P P_3 + \beta_P P_3^3 + \gamma_P P_3^5 - g_{33mn}\frac{\partial^2 P_3}{\partial x_m \partial x_n} - 2Q_{kli3}\sigma_{kl}P_3 = E_3, \qquad (1)$$

where the coefficient $\alpha_P(T) = \alpha_P^{(T)}(T - T_C)$, T is the absolute temperature, $T_C$ is the Curie temperature of the paraelectric-to-ferroelectric phase transition. The parameters $\beta_P$, and $\gamma_P$ are coefficients of LGD potential expansion on the polarization powers. $\sigma_{kl}$, and $Q_{ijkl}$ are respectively elastic stress and electrostriction stress tensor. Flexoelectric effect is regarded as small. Boundary conditions for the polarization $P_3$ at the particle surface S are regarded to be natural, $(\partial P_3/\partial \mathbf{n})|_S = 0$.

Electric field $E_i$ is defined via electric potential as $E_i = -\partial \varphi / \partial x_i$. For a ferroelectric-dielectric, the electric potential $\varphi$ can be found self-consistently from the Laplace equation outside the nanoparticle $\varepsilon_0 \varepsilon^e \Delta \varphi_e = 0$ and Poisson equation inside it

$$\varepsilon_0 \varepsilon_{ij}^b \frac{\partial^2 \varphi}{\partial x_i \partial x_j} = \frac{\partial P_k}{\partial x_k}, \qquad (2)$$

$\varepsilon_0 = 8.85 \times 10^{-12}$ F/m the dielectric permittivity of vacuum, $\varepsilon_{ij}^b$ is background permittivity [66]. Free charges are regarded absent inside the particle.

Corresponding electric boundary conditions of potential continuity at the mechanically-free particle surface S, $(\varphi_e - \varphi_i)|_S = 0$. The boundary condition for the normal components of electric displacements should take into account the surface screening produced by e. g. ambient free charges at the particle surface S, $\left((\mathbf{D}_e - \mathbf{D}_i)\mathbf{n} + \varepsilon_0 \frac{\varphi_i}{\lambda}\right)\bigg|_S = 0$, where $\lambda$ is the surface screening length. The potential is constant at the particle-electrode interface, i.e. $\varphi_i|_{z=0} = 0$. Surface screening leads to the decrease of external field inside the particle, as well as to the decrease of bare depolarization field caused by the polarization gradient.

Within a phenomenological approach, linear and biquadratic ME couplings contribution to the system free energy are described by the terms $\mu_{ij} P_i M_j$ and $\xi_{ijkl} P_i P_j M_k M_l$, respectively (**P** is polarization and **M** is magnetization, and $\mu_{ij}$ and $\xi_{ijkl}$ are corresponding tensors of ME effects, respectively) [67, 68, 69, 70, 71]. The PME coupling contribution is described by the term $\eta_{ijk} P_i M_j M_k$ [67, 72]. Let us use the phenomenological LGD-based model for the PME coefficients calculation [72, 73]. Assuming that the magnetization M is linearly proportional to the applied magnetic field H, $M \approx \chi_{FM}(T) H$, the PME coefficient η has the form [72]:

$$\eta(T) = -P_S(T) \chi_{FE}(T) (\chi_M(T))^2 \xi_{MP} . \quad (3)$$

Here the spontaneous polarization $P_S(T)$ is the averaged over the particle volume spontaneous polarization $\langle P_3(\mathbf{r}) \rangle$ calculated from Eq.(1) at H=0 and E=0. Functions $\chi_M(T)$ and $\chi_{FE}(T)$ are averaged over the particle volume linear magnetic susceptibility and dielectric susceptibility in the ferroelectric phase, respectively. Ferroelectric susceptibility can be calculated from Eq.(1) using the definition

$$\chi_{FE}(T) = \left. \frac{\partial \langle P_3 \rangle}{\partial E_3} \right|_{E_3=0} . \quad (4)$$

Approximate expression for magnetic susceptibility is taken the same as in Ref.[72]:

$$\chi_M(T) = \frac{\mu_0}{\alpha_M^{(T)}(T - \theta) + \xi_{LM} L^2 + \xi_{MP} P_S^2(T)} . \quad (5)$$

Equations (3) and (4) are valid in the ferroelectric-antiferromagnetic phase (with nonzero antiferromagnetic long-range order parameter $L \neq 0$) as well as in the ferroelectric – paramagnetic phase without any magnetic order $M$ =L=0. Parameters $\xi_{LM}$ and $\xi_{MP}$ the biquadratic magnetoelectric coefficients that couple polarization and magnetic order parameters in the magnetoelectric energy $G_{ME} = \frac{1}{2}(\xi_{MP} M^2 + \xi_{LP} L^2) P^2$. It should be noted, that only two coefficients in magnetic energy, $G_M = \frac{\alpha_L(T)}{2} L^2 + \frac{\beta_L}{4} L^4 + \frac{\alpha_M(T)}{2} M^2 + \frac{\beta_M}{4} M^4 - \mu_0 M H + \frac{\xi_{LM}}{2} L^2 M^2$, are assumed to be dependent on temperature, namely $\alpha_L(T) = \alpha_M^{(T)}(T - T_N)$ and $\alpha_M(T) = \alpha_M^{(T)}(T - \theta)$ where θ is the magnetic Curie temperature, and $T_N$ is the Neel temperature.

## 2.2. Analytical approximation based on finite element modeling

Using finite element modeling (FEM) we numerically calculated the spatial distribution [**Fig.2(a)**] and average electric field inside the particles. Ferroelectric parameters of BiFeO$_3$ are listed in **Table I**.

**Table I**. Parameters of BiFeO$_3$ used in our calculations

| Parameter | SI units | Value for BiFeO$_3$ |
|---|---|---|
| Spontaneous polarization $P_S$ | m/C$^2$ | 1 |
| Electrostriction coefficient $Q_{12}$ | m$^4$/C$^2$ | −0.05 |
| Electrostriction coefficient $Q_{11}$ | m$^4$/C$^2$ | −0.1 |
| Background permittivity $\varepsilon_b$ | dimensionless | 10 |
| Ambient permittivity $\varepsilon_e$ | dimensionless | 1 |
| Gradient coefficient $g_{11}$ | m$^3$/F | $10^{-10}$ |
| LGD coefficient $\alpha_S$ | m$^2$/F | $10^{-4}$ |
| LGD coefficient $\beta$ | J m$^5$/C$^4$ | $10^7$ |
| LGD coefficient $\alpha$ | m/F | $-10^7$ (at 300 K) |
| Ferroelectric Curie temperature $T_c$ | K | 1100 |
| Temperature coefficient $\alpha_T$ | m/(K F) | $0.9 \times 10^6$ |
| Antiferromagnetic Neel temperature | K | 650 |
| Surface screening length $\lambda$ | nm | $10^{-3}$ to $10^2$ |
| Universal dielectric constant $\varepsilon_0$ | F/m | $8.85 \times 10^{-12}$ |

Numerical results were approximated analytically. Obtained electric field dependence on the surface screening length $\lambda$ has the following form:

$$E_{dX} \approx -\frac{P_X}{\varepsilon_0} \frac{\lambda n_\infty(a,b,c)}{\lambda + R n_\infty(a,b,c)} \qquad (6)$$

Here $n_\infty$ is the "bare" depolarization factor of the system without screening charges (the case of the limit $\lambda \to \infty$). $R$ is the characteristic length-scale, proportional to the size $a$ of island along the polar axis. Using Eq.(6) we can introduce the effective depolarization factor ($n_d(a,b,c) = -\varepsilon_0 \frac{E_{dX}}{P_X}$) dependent on the semi-ellipsoid geometry as

$$n_d(a,b,c) = \frac{\lambda n_\infty(a,b,c)}{\lambda + R n_\infty(a,b,c)} \qquad (7)$$

High accuracy of approximation (7) becomes evident from **Fig. 2(b)**.

The fitting with Eq.(7) allows us to obtain parameters $n_\infty$ and $R$ for a set of island sizes $a$, $b$ and $c$, which are the lengths of ellipsoid semi-axis. These sets were fitted with the Pade-approximations of the following form:

$$n_\infty(a,b,c) \approx \frac{b}{\varepsilon_b b + \varepsilon_e a}\left(\frac{c^2}{c^2 + 0.7ac + a^2\dfrac{b}{b+0.075a}}\right), \tag{8}$$

$$R(a,b,c) \approx a\left(0.62 + 0.19\frac{a}{b} + 0.25\frac{a}{c}\right). \tag{9}$$

Note that the pre-factor $\dfrac{b}{\varepsilon_b b + \varepsilon_e a}$ in Eq.(8) is the exact expression for depolarization factor of elliptical cylinder with semi-axes *a* and *b*. High accuracy of approximations (8)-(9) becomes evident from **Fig. 2(b)-(d)**.

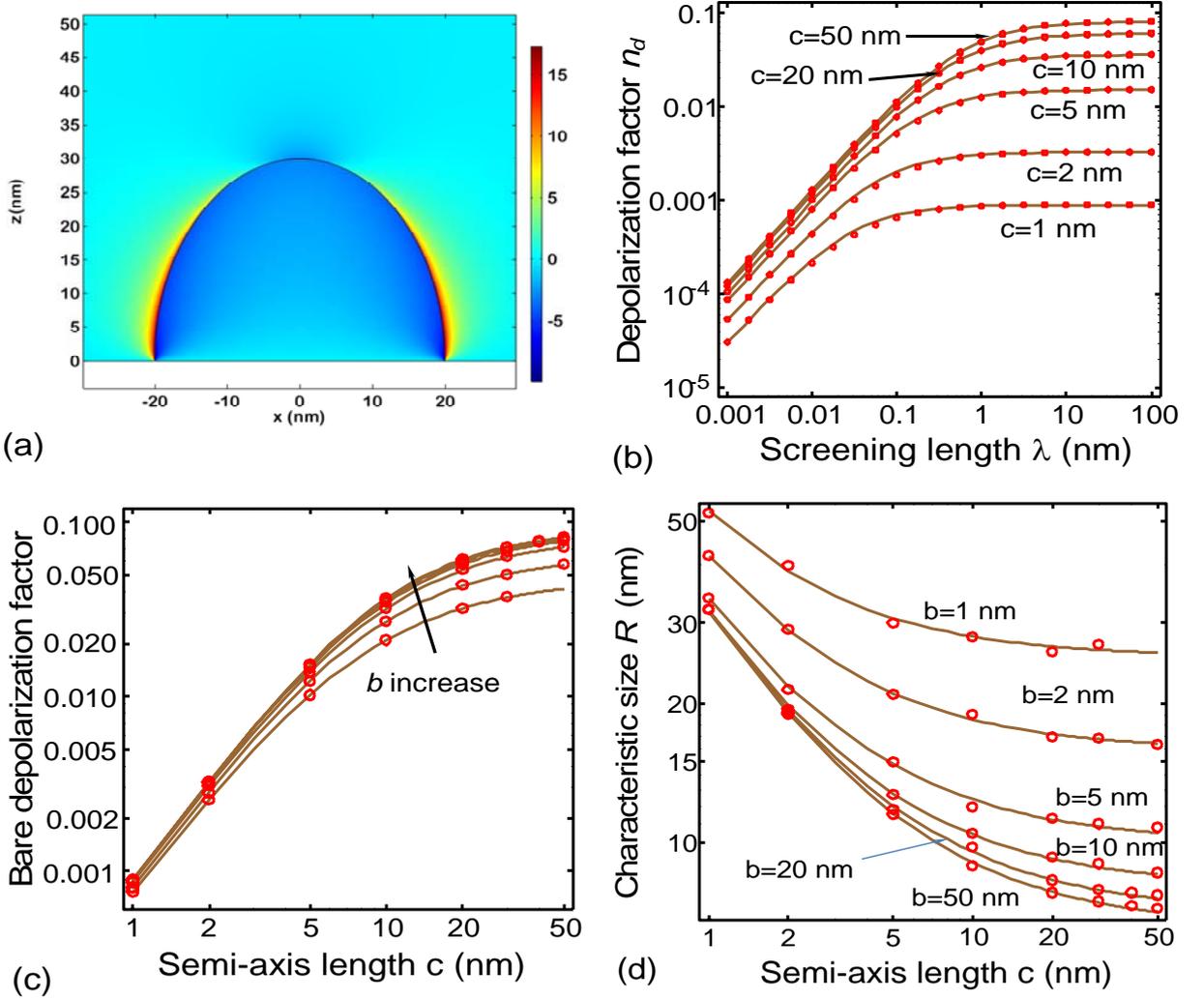

**FIG. 2. (a)** Distribution of the electric field component $E_x$ in the system calculated numerically. **(b)** Dependence of depolarization factor $n_X$ on the screening length for different values of semi-axis length *c* (numbers near the curves) and fixed value of *a*=10 nm *b*=20 nm. Dependences of parameters $n_\infty$ **(c)** and *R* **(d)** on the semi-axis length *c* for fixed values of *a*=10 nm and *b*=1, 2, 5, 10, 20, 50 nm (numbers near the curves). Relative dielectric permittivities $\varepsilon_b = 10$ and $\varepsilon_e = 1$. Symbols are results of FEM calculations, solid curves represent fitting with Eq.(6) - (9), respectively.

Allowing for expressions (6)-(9), the transition temperature to paraelectric phase $T_{cr}(a,b,c)$ can be defined from the condition $\alpha + \dfrac{n_d}{\varepsilon_0} = 0$ and given by analytical expression

$$T_{cr}(a,b,c) = T_C - \frac{n_d(a,b,c)}{\alpha_T \varepsilon_0}. \tag{10}$$

Equation (10) allows to write analytical expressions for the average spontaneous polarization, and linear dielectric susceptibility by conventional form

$$P_S = \begin{cases} \sqrt{\dfrac{\alpha_T}{\beta}(T_{cr}(a,b,c)-T)}, & T < T_{cr}, \\ 0, & T > T_{cr}. \end{cases} \tag{11}$$

$$\chi_{FE}(T) = \begin{cases} \dfrac{1}{2\alpha_T(T_{cr}(a,b,c)-T)}, & T < T_{cr}, \\ \dfrac{1}{\alpha_T(T - T_{cr}(a,b,c))}, & T > T_{cr}. \end{cases} \tag{12}$$

Allowing for Eqs.(3), (5), (11) and (12) the analytical expression for PME coefficient acquires the form:

$$\eta(T) = \begin{cases} \dfrac{-\xi_{MP}(\chi_M(T))^2}{2\sqrt{\alpha_T \beta(T_{cr}(a,b,c)-T)}}, & T < T_{cr}, \\ 0, & T > T_{cr}. \end{cases} \tag{13}$$

It follows from the obtained formula (10) – (13), that the main peculiarities of the ellipsoidal nanoparticles originate from depolarization field contribution.

### 3. RESULTS AND DISCUSSION
**3.1. Size effects of phase diagrams and average polarization**

Phase diagrams of semi-ellipsoidal BiFeO$_3$ nanoparticles in coordinates relative temperature $T/T_C$ - length of the particle longer semi-axis $a$ are shown in **Figs.3(a)** and **3(b)** ($T_C$ is bulk Curie temperature). The boundary between paraelectric (PE) and ferroelectric (FE) phases (that is in fact the critical temperature of the size-induced phase transition $T_{cr}(a,b,c)$) depends on the semi-ellipsoid sizes a, b and c (multiple size effect). The size effect manifested itself in the ferroelectricity disappearance at the critical size $a_{cr}(b,c)$ for which $T_{cr} = 0$ followed, by the monotonic increase of the transition temperature with the size $a$ increase and its further saturation to $T_C$ for the sizes $a \gg 100$ nm. Different curves are calculated for several values of semi-axis $b =$ 3, 10, 30 and 100 nm and fixed particle height $c$. **Figure 3(a)** corresponds to the particles of small height $c=10$ nm, and **Fig.3(b)** for $c=100$ nm. The critical size $a_{cr}(b,c)$ monotonically decreases as well as the phase boundary between PE and FE phase shifts to the right with $b$ increase at the same $c$ values [compare different curves in **Figs.3(a)** and **3(b)**]. Moreover, the critical sizes $a_{cr}(b,c)$ calculated at $c=10$ nm are essentially smaller than the sizes calculated at $c=100$ nm at the same $b$-values [compare the curves calculated in **Figs.3(a)** and **3(b)**]. In numbers, the PE-FE transition exists for all values of chosen sizes. At $c = 10$ nm, the critical size $a_{cr}(b,c)$ varies in the narrow range (10 – 12) nm, and the curves calculated for different $b$ values are very close to one another. At $c = 100$ nm, the critical size $a_{cr}(b,c)$ varies in the wider range (15 – 45) nm, and the curves calculated for different $b$ values are well-separated from one another.

Hence the analysis of **Figs. 3(a)** and **3(b)** allows us to conclude that the size effect of the phase diagrams is sensitive to the value of the particle aspect ratio in the polarization direction, $bc/a^2$, and less sensitive to the absolute values of the sizes per se. The smaller is the ratio, the smaller is the depolarization field and hence the higher is the transition temperature and the smaller is the critical size. The result seems nontrivial a priory.

The spontaneous polarization dependence on the length of ellipsoid semi-axis $a$ calculated for different values of semi-axis $c = 10$ nm and 100 nm and room temperature are shown in **Figs. 3(c)** and **3(d)**, respectively**.** The values of another semi-axis $b$ are chosen the same as in **Figs. 3(a)** and **3(b)** [see different curves calculated for $b=$ 3, 10, 30 and 100 nm]. The polarization curves calculated for different $b$ values are very close to one another at $c = 10$ nm, and are well-separated from one another at $c = 100$ nm. The spontaneous polarization appears at the critical size $a_{cr}(b,c)$ and increases with the size $a$ increase. The polarization saturates to the bulk value ~ 1 C/m$^2$ at sizes $a \gg 100$ nm. Note that the polarization of the particles with height $c = 10$ nm saturates essentially faster than the one for the particles with $c = 100$ nm [compare curves saturation in **Figs. 3(c)** and **3(d)**].

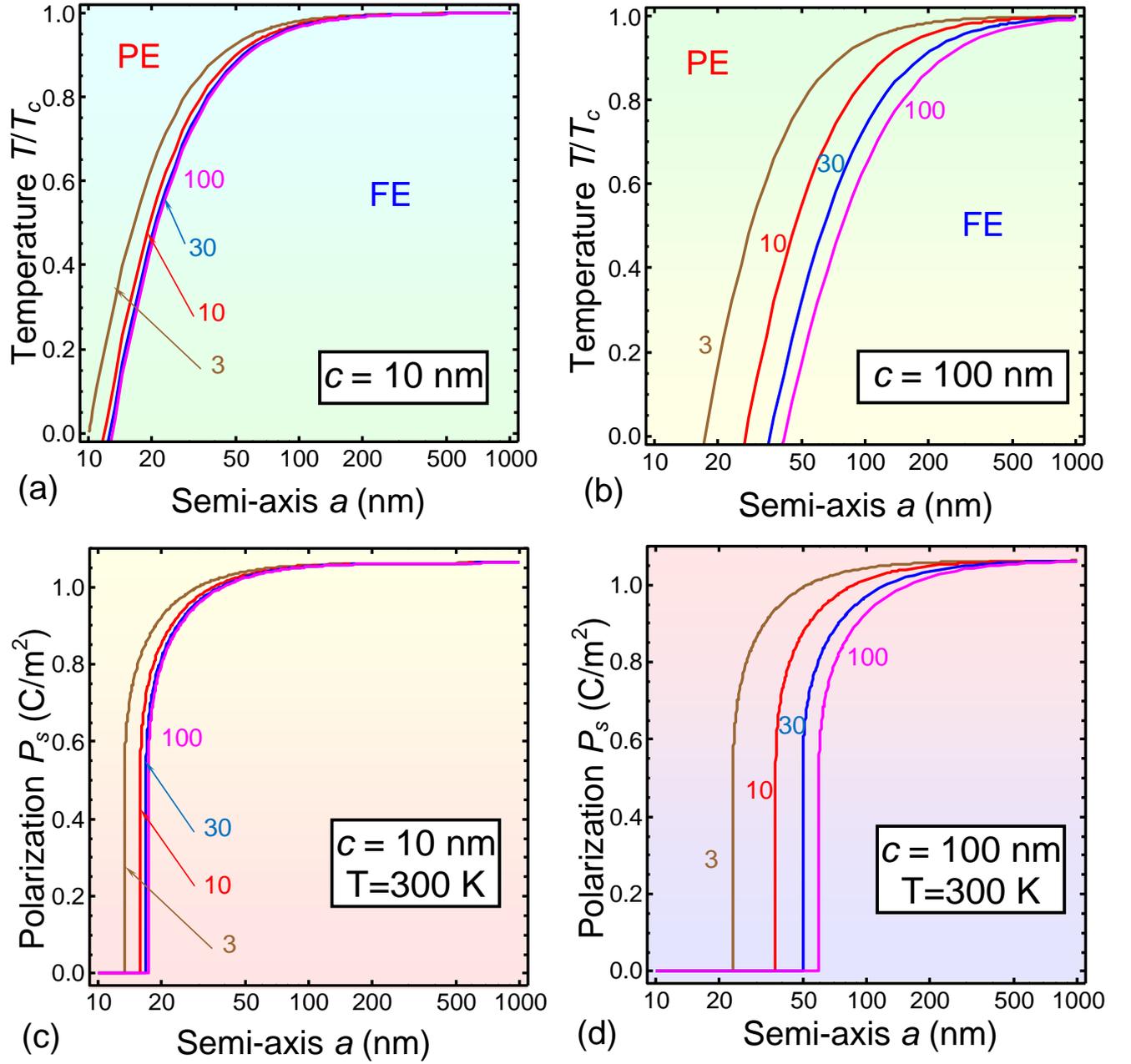

**FIG. 3.** Phase diagrams in coordinates temperature – length of ellipsoid semi-axis *a* calculated for different values of semi-axis *c* = 10 nm and 100 nm (panels **(a)** and **(b)** respectively) and axis *b*= 3, 10, 30 and 100 nm (see numbers near the curves). The spontaneous polarization dependence on the length of ellipsoid semi-axis *a* calculated at room temperature for different values of semi-axis *c*= 10 nm and 100 nm (panels **(c)** and **(d)** respectively) and axis *b*= 3, 10, 30 and 100 nm (see numbers near the curves). Screening length λ=1 nm, other parameters corresponding to BiFeO$_3$ compound are listed in **Table I**.

### 3.2. Size effect of paramagnetoelectric coefficient

The dependences of PME effect coefficient on the length of ellipsoid semi-axis *a* calculated at room temperature for different values of semi-axis *c* = 10 nm and 100 nm are shown in **Figs. 4(a)** and **4(b)**, respectively. The values of another semi-axis *b* are chosen the same as in the previous figures [see different curves calculated for *b*= 3, 10, 30 and 100 nm]. PME coefficient is normalized on its bulk value. The PME coefficient is zero at sizes $a < a_{cr}(b,c)$ because of spontaneous polarization disappearance, it appears at $a < a_{cr}$ and diverges at the critical size $a = a_{cr}(b,c)$, and then it decreases with the size *a* increase. The PME coefficient saturates to the bulk value at sizes $a \gg 100$ nm. The divergences at $a = a_{cr}(b,c)$ demonstrate the possibility to obtain giant PME effect in BiFeO$_3$ nanoparticles in the vicinity of size-induced transition from the FE phase to a PE phase. In particular the normalized PME coefficient is essentially higher than unity for sizes $a_{cr}(b,c) \le a < 2a_{cr}(b,c)$. The behavior of PME coefficient reproduces the behavior of the dielectric susceptibility given by Eq.(12) in the framework of our model. Note that the PME coefficient of the particles with height *c* = 10 nm saturates essentially faster than the one for the particles with *c* = 100 nm [compare curves saturation in **Figs. 4(a)** and **4(b)**]. The PME coefficient curves calculated for different *b* values are very close to one another at *c* = 10 nm, and are well-separated from one another at *c* = 100 nm.

The comparative analyses of the **Figs.3(c)-(d)** and **Figs.4(a)-(b)** approves our conclusion that the size effect of the spontaneous polarization and PME coefficient is sensitive to the particle aspect ratio in the polarization direction, $bc/a^2$, and less sensitive to the absolute values of the sizes.

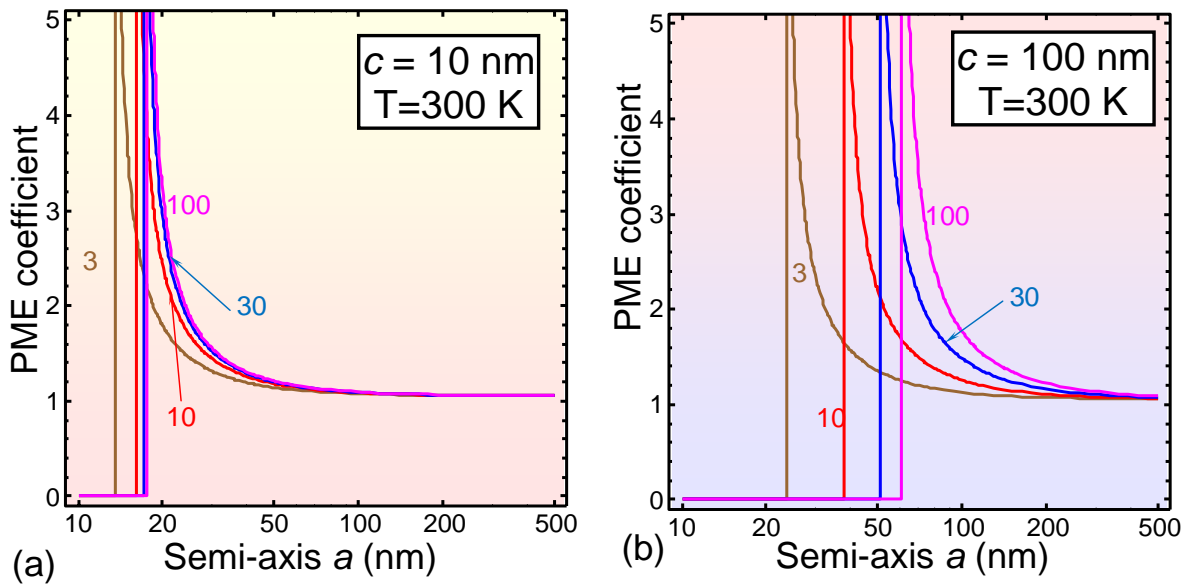

**FIG. 4.** Normalized paramagnetoelectric (PME) effect coefficient dependence on length of ellipsoid semi-axis *a* calculated for different values of semi-axis *c*= 10 nm and 100 nm (panels **(a)** and **(b)**, respectively). Different curves at each plot correspond to the different value of the axis *b*= 3, 10, 30 and 100 nm (indicated by numbers near the curves). Screening length λ=1 nm, room temperature, other parameters corresponding to BiFeO$_3$ compound are listed in **Table I**.

## 4. CONCLUSIONS

We have studied the size effects on the phase diagrams, ferroelectric and magnetoelectric properties of semi-ellipsoidal BiFeO$_3$ nanoparticles clamped to a rigid conductive substrate. The spatial distribution of the spontaneous polarization vector inside the ferroelectric nanoparticles, phase diagrams and paramagnetoelectric coefficient were calculated in the framework of the Landau-Ginzburg-Devonshire approach, classical electrostatics and elasticity theory. Rather rigorous analytical expressions were derived for the dependences of the ferroelectric transition temperature, average polarization, linear dielectric susceptibility and paramagnetoelectric coefficient on the particle sizes for a general case of a semi-ellipsoidal nanoparticles with three different semi-axes *a*, *b* and height *c*. Due to the essential decrease of depolarization field the in-plane orientation of the spontaneous polarization along the longer semi-ellipsoidal axis *a* is energetically preferable for the nanoparticles of small height *c<a*. The analyses of the obtained results leads to the conclusion that the size effect on the phase diagrams, spontaneous polarization and paramagnetoelectric coefficient is rather sensitive to the particle sizes aspect ratio in the polarization direction, $bc/a^2$, and less sensitive to the absolute values of the sizes per se. This opens the way to govern the properties by the choice of this ratio value.

**ASKNOWLEDGMENTS.** M.V.S. and D.V.K acknowledge MK-1720.2017.8, BRFFI (# F16R-066) RFFI (# 16-58-00082). E.A.E. and A.N.M. acknowledge the Center for Nanophase Materials Sciences, which is a DOE Office of Science User Facility, CNMS2016-061.

**Authors contribution.** A.N.M. generated the research idea, states the mathematical problem and wrote the manuscript draft. V.V.K. with E.A.E. assistance performed numerical simulations. E.A.E. and A.N.M. derived analytical expressions. M.D.G. worked on the results interpretation and manuscript improvement. M.V.S. and D.A.K. valuably contributed to the research idea and manuscript improvement.

## REFERENCES

[1] M. Fiebig, Revival of the magnetoelectric effect, Journal of Physics D: Applied Physics, 38 (2005) 123-152.

[2] N. A. Spaldin and M. Fiebig, Materials science. The renaissance of magnetoelectric multiferroics, Science, 309 (2005) 391-392.

[3] J.M. Rondinelli, N.A. Spaldin, Structure and properties of functional oxide thin films: Insights from electronic-structure calculations, Advanced Materials, 23 (2011) 3363–3381.

[4] A. P. Pyatakov, A. K.Zvezdin, Magnetoelectric and multiferroic media, Physics-Uspekhi, 55:6 (2012) 557-581.

[5] J. F. Scott, Data storage: Multiferroic memories, Nature materials, 6 (2007) 256-257.

[6] R. Ramesh & A. Nicola Spaldin, Multiferroics: progress and prospects in thin films, Nature Mater, 6 (2007) 21-29.

[7] P.J. Ryan, J-W. Kim, T. Birol, P. Thompson, J-H. Lee, X. Ke, P.S. Normile, E. Karapetrova, P. Schiffer, S.D. Brown, C.J. Fennie, D.G. Schlom, Reversible control of magnetic interactions by electric field in a single-phase material, Nat Commun, 4 (2013) 1334.

[8] M.J. Haun, E. Furman, T.R. Halemane, L.E. Cross, Thermodynamic theory of the lead zirconate-titanate solid solution system, part IV: tilting of the oxygen octahedral, Ferroelectrics, 99 (1989) 55-62.

[9] E.V. Balashova, A.K. Tagantsev, Polarization response of crystals with structural and ferroelectric instabilities, Phys Rev B, 48 (1993) 9979-9986.

[10] A.K. Tagantsev, E. Courtens, L. Arzel, Prediction of a low-temperature ferroelectric instability in antiphase domain boundaries of strontium titanate, Phys Rev B, 64 (2001) 224107.

[11] S.L. Hou, N. Bloembergen, Paramagnetoelectric Effects in $NiSO_4 \cdot 6 H_2O$, Phys Rev, 138 (1965) 1218-1226.

[12] V.V. Shvartsman, S. Bedanta, P. Borisov, W. Kleemann, $(Sr,Mn)TiO_3$: A Magnetoelectric Multiglass, Phys Rev Lett, 101(2008) 165704.

[13] B. Howes, M. Pelizzone, P. Fischer, C. Tabaresmunoz, J-P. Rivera, H. Schmid, Characterisation of some magnetic and magnetoelectric properties of ferroelectric $Pb(Fe_{1/2}Nb_{1/2})O_3$, Ferroelectrics, 54 (1984) 317-320.

[14] T. Watanabe, K. Kohn, Magnetoelectric effect and low temperature transition of $PbFe_{0.5}Nb_{0.5}O_3$ single crystal, Phase Transitions, 15 (1989) 57-68.

[15] W. Kleemann, V.V. Shvartsman, P. Borisov, A. Kania, Coexistence of antiferromagnetic and spin cluster glass order in the magnetoelectric relaxor multiferroic $PbFe_{0.5}Nb_{0.5}O_3$, Phys Rev Lett, 105 (2010) 257202.

[16] V. V. Laguta, A. N. Morozovska, E. A. Eliseev, I. P. Raevski, S. I. Raevskaya, E. I. Sitalo, S. A. Prosandeev, and L. Bellaiche, Room-temperature paramagnetoelectric effect in magnetoelectric multiferroics $Pb(Fe_{1/2}Nb_{1/2})O_3$ and its solid solution with $PbTiO_3$, Journal of Materials Science, 51 (2016) 5330-5342.

[17] J. Seidel, L.W. Martin, Q. He, Q. Zhan, Y.-H. Chu, A. Rother, M. E. Hawkridge, P. Maksymovych, P. Yu, M. Gajek, N. Balke, S. V. Kalinin, S. Gemming, F. Wang, G. Catalan, J. F. Scott, N. A. Spaldin, J. Orenstein and R. Ramesh, Conduction at domain walls in oxide multiferroics, Nature Materials, 8 (2009) 229 – 234.

[18] J. Seidel, P. Maksymovych, Y. Batra, A. Katan, S.-Y. Yang, Q. He, A. P. Baddorf, S. V. Kalinin, C.-H.Yang, J.-C. Yang, Y.-H. Chu, E. K. H. Salje, H.Wormeester, M. Salmeron and R. Ramesh, Domain Wall Conductivity in La-Doped $BiFeO_3$, Phys. Rev. Lett., 105 (2010) 197603.

[19] Q. He, C.-H. Yeh, J.-C. Yang, G. Singh-Bhalla, C.-W. Liang, P.-W. Chiu, G. Catalan, L.W. Martin, Y.-H. Chu, J. F. Scott, and R. Ramesh, Magnetotransport at Domain Walls in BiFeO$_3$, Phys. Rev. Lett., 108 (2012) 067203.

[20] G. Catalan, J. Seidel, R. Ramesh, and J. F. Scott, Domain wall nanoelectronics, Rev. Mod. Phys., 84 (2012) 119.

[21] R. K. Vasudevan, A. N. Morozovska, E. A. Eliseev, J. Britson, J.-C. Yang, Y.-H. Chu, P. Maksymovych, L. Q. Chen, V. Nagarajan, S. V. Kalinin. Domain wall geometry controls conduction in ferroelectrics. Nano Letters, 12 (11) (2012) 5524–5531.

[22] Anna N. Morozovska, Rama K. Vasudevan, Peter Maksymovych, Sergei V. Kalinin and Eugene A. Eliseev. Anisotropic conductivity of uncharged domain walls in BiFeO$_3$, Phys. Rev. B., 86 (2012) 085315.

[23] P. Fischer, M. Polomska, I. Sosnowska, M. Szymanski, Temperature dependence of the crystal and magnetic structures of BiFeO$_3$, J. Phys. C: Solid St. Phys., 13 (1980) 1931-1940.

[24] Gustau Catalan, James F. Scott, Physics and Applications of Bismuth Ferrite, Adv. Mater., 21 (2009) 1–23.

[25] J. Wang, J. B. Neaton, H. Zheng, V. Nagarajan, S. B. Ogale, B. Liu, D. Viehland, V. Vaithyanathan, D. G. Schlom, U. V. Waghmare, N. A. Spaldin, K. M. Rabe, M. Wuttig, R. Ramesh Science, Epitaxial BiFeO$_3$ Multiferroic Thin Film Heterostructures, 299 (2003) 1719.

[26] Y.-H. Chu, Qian Zhan, Lane W. Martin, Maria P. Cruz, Pei-Ling Yang, Gary W. Pabst, Florin Zavaliche, Seung-Yeul Yang, Jing-Xian Zhang, Long-Qing Chen, Darrell G. Schlom, I.-Nan Lin, Tai-Bor Wu, and Ramamoorthy Ramesh, Nanoscale Domain Control in Multiferroic BiFeO$_3$ Thin Films, Adv. Mater., 18 (2006) 2307.

[27] Ying-Hao Chu, Lane W. Martin, Mikel B. Holcomb, Martin Gajek, Shu-Jen Han, Qing He, Nina Balke, Chan-Ho Yang, Donkoun Lee, Wei Hu, Qian Zhan, Pei-Ling Yang, Arantxa Fraile-Rodríguez, Andreas Scholl, Shan X. Wang & R. Ramesh, Electric-field control of local ferromagnetism using a magnetoelectric multiferroic, Nature Materials, 7 (2008) 478 – 482.

[28] Peter Maksymovych, Mark Huijben, Minghu Pan, Stephen Jesse, Nina Balke, Ying-Hao Chu, Hye Jung Chang, Albina Y. Borisevich, Arthur P. Baddorf, Guus Rijnders, Dave H. A. Blank, Ramamoorthy Ramesh, and Sergei V. Kalinin, Ultrathin limit and dead-layer effects in local polarization switching of BiFeO$_3$, Phys. Rev. B, 85 (2012) 014119.

[29] C. Beekman, W. Siemons, M. Chi, N. Balke, J. Y. Howe, T. Z. Ward, P. Maksymovych, J. D. Budai, J. Z. Tischler, R. Xu, W. Liu, and H. M. Christen., Ferroelectric Self-Poling, Switching, and Monoclinic Domain Configuration in BiFeO$_3$,Thin Films, Adv. Funct. Mater., 26 (2016) 5166–5173.

[30] G. A. Smolenskii, L. I. Chupis, Ferroelectromanetgs, Sov. Phys. Usp., 25 (1982) 475.

[31] A.Y. Borisevich, O.S. Ovchinnikov, Hye Jung Chang, M.P. Oxley, Pu Yu, Jan Seidel, E.A. Eliseev, A.N. Morozovska, Ramamoorthy Ramesh, S.J. Pennycook, S.V. Kalinin., Beyond Condensed Matter Physics on the Nanoscale: The Role of Ionic and Electrochemical Phenomena in the Physical Functionalities of Oxide Materials,ACS Nano, 4 (2010) 6071–6079.

[32] Nina Balke, Benjamin Winchester, Wei Ren, Ying Hao Chu, Anna N. Morozovska, Eugene A. Eliseev, Mark Huijben, Rama K. Vasudevan, Petro Maksymovych, Jason Britson, Stephen Jesse, Igor Kornev, Ramamoorthy Ramesh, Laurent Bellaiche, Long Qing Chen, and Sergei V. Kalinin, Enhanced electric conductivity at ferroelectric vortex cores in $BiFeO_3$, Nature Physics, 8 (2012) 81–88.

[33] Y.-M. Kim, A. Kumar, A. Hatt, A. N. Morozovska, A. Tselev, M. D. Biegalski, I. Ivanov, E. A. Eliseev, S. J. Pennycook, J. M. Rondinelli, S. V. Kalinin, A. Y. Borisevich, Interplay of octahedral tilts and polar order in $BiFeO_3$ films, Adv. Mater., 25 (2013) 2497–2504.

[34] R. K. Vasudevan, W. Wu, J. R. Guest, A. P. Baddorf, A. N. Morozovska, E. A. Eliseev, N. Balke, V. Nagarajan, P. Maksymovych, Domain Wall Conduction and Polarization-Mediated Transport in Ferroelectrics, Adv. Funct. Mater., 23 (2013) 2592–2616.

[35] Young-Min Kim, Anna Morozovska, Eugene Eliseev, Mark Oxley, Rohan Mishra, Tor Grande, Sverre Selbach, Sokrates Pantelides, Sergei Kalinin, and Albina Borisevich, Direct observation of ferroelectric field effect and vacancy-controlled screening at the BiFeO3-LaxSr1-xMnO3 interface, Nature Materials, 13 (2014) 1019–1025.

[36] B. Winchester, N. Balke, X. X. Cheng, A. N. Morozovska, S. Kalinin, and L. Q. Chen, Electroelastic fields in artificially created vortex cores in epitaxial $BiFeO_3$ thin films, Applied Physics Letters, 107 (2015) 052903.

[37] James F. Scott, Advanced Materials, Iso-Structural Phase Transitions in $BiFeO_3$, 22 (2010) 2106-2107.

[38] Samar Layek and H. C. Verma, Magnetic and dielectric properties of multiferroic BiFeO3 nanoparticles synthesized by a novel citrate combustion method, Adv. Mat. Lett., 3 (2012) 533-538.

[39] Fengzhen Huang, Zhijun Wang, Xiaomei Lu, Junting Zhang, Kangli Min, Weiwei Lin, Ruixia Ti, TingTing Xu, Ju He, Chen Yue & Jinsong Zhu, Magnetism of $BiFeO_3$ nanoparticles Peculiar with size approaching the period of the spiral spin structure, Scientific Reports, 3 (2013) 2907.

[40] D. Yadlovker, S. Berger, Uniform orientation and size of ferroelectric domains, Phys. Rev. B., 71 (2005) 184112.

[41] D. Yadlovker, S. Berger, Reversible electric field induced nonferroelectric to ferroelectric phase transition in single crystal nanorods of potassium nitrate, Appl. Phys. Lett., 91 (2007) 173104.

[42] D. Yadlovker, S. Berger, Nucleation and growth of single crystals with uniform crystallographic orientation inside alumina nanopores, J. Appl. Phys., 101 (2007) 034304.

[43] M. H. Frey, D. A. Payne, Grain-size effect on structure and phase transformations for barium titanate, Phys. Rev. B, 54 (1996) 3158- 3168.

[44] Z. Zhao, V. Buscaglia, M. Viviani, M.T. Buscaglia, L. Mitoseriu, A. Testino, M. Nygren, M. Johnsson, P. Nanni, Grain-size effects on the ferroelectric behavior of dense nanocrystalline $BaTiO_3$ ceramics, Phys. Rev. B, 70 (2004) 024107.

[45] E. Erdem, H.-Ch. Semmelhack, R. Bottcher, H. Rumpf, J. Banys, A.Matthes, H.-J. Glasel, D. Hirsch, E. Hartmann, Study of the tetragonal-to-cubic phase transition in $PbTiO_3$ nanopowders, J. Phys.: Condens. Matter, 18 (2006) 3861–3874.

[46] I.S. Golovina, S.P. Kolesnik, V. Bryksa, V.V. Strelchuk, I.B. Yanchuk, I.N. Geifman, S.A. Khainakov, S.V. Svechnikov, A.N. Morozovska, Defect driven ferroelectricity and magnetism in nanocrystalline $KTaO_3$, Physica B: Condensed Matter., 407 (2012) 614-623.

[47] I.S. Golovina, V.P. Bryksa, V.V. Strelchuk, I.N. Geifman, A.A. Andriiko, Size effects in the temperatures of phase transitions in $KNbO_3$ nanopowder, J. Appl. Phys., 113 (2013) 144103.

[48] I.S. Golovina, V.P. Bryksa, V.V. Strelchuk, I.N. Geifman, Phase transitions in the nanopowders KTa0.5Nb0.5O3 studied by Raman spectroscopy, Functional Materials., 20 (2013) 75-80.

[49] I.S. Golovina, B.D. Shanina, S.P. Kolesnik, I. N. Geifman, A. A. Andriiko, Magnetic properties of nanocrystalline $KNbO_3$, J. Appl. Phys., 114 (2013)174106.

[50] T. Yu, Z. X. Shen, W. S. Toh, J. M. Xue, J. Wang, Size effect on the ferroelectric phase transition in $SrBi_2Ta_2O_9$ nanoparticles, J. Appl. Phys., 94 (2003) 618.

[51] H. Ke, D. C. Jia, W. Wang, Y. Zhou, Ferroelectric phase transition investigated by thermal analysis and Raman scattering in $SrBi_2Ta_2O_9$ nanoparticles, Solid State Phenomena Vols., 121-123 (2007) 843-846.

[52] P. Perriat, J. C. Niepce, G. Caboche, Thermodynamic considerations of the grain size dependency of material properties: a new approach to explain the variation of the dielectric permittivity of $BaTiO_3$ with grain size, Journal of Thermal Analysis and Calorimetry, 41 (1994) 635-649.

[53] H. Huang, C. Q. Sun, P. Hing, Surface bond contraction and its effect on the nanometric sized lead zirconate titanate, J. Phys.: Condens. Matter, 12 (2000) 127–132.

[54] H. Huang, C. Q. Sun, Z. Tianshu, P. Hing, Grain-size effect on ferroelectric $Pb(Zr_{1-x}Ti_x)O_3$ solid solutions induced by surface bond contraction, Phys. Rev. B, 63 (2001) 184112.

[55] M. Wenhui, Surface tension and Curie temperature in ferroelectric nanowires and nanodots, Appl. Phys. A, 96 (2009) 915–920.

[56] E.A. Eliseev, A.N. Morozovska, M.D. Glinchuk, R. Blinc, Spontaneous flexoelectric/flexomagnetic effect in nanoferroics, Phys. Rev. B., 79 (2009) 165433.

[57] A. N. Morozovska, E. A. Eliseev, M.D. Glinchuk, Ferroelectricity enhancement in confined nanorods: Direct variational method, Phys. Rev. B, 73 (2006) 214106.

[58] A. N. Morozovska, M. D. Glinchuk, E.A. Eliseev, Phase transitions induced by confinement of ferroic nanoparticles, Phys. Rev. B, 76 (2007) 014102.

[59] A.N. Morozovska, I.S. Golovina, S.V. Lemishko, A.A. Andriiko, S.A. Khainakov, E.A. Eliseev, Effect of Vegard strains on the extrinsic size effects in ferroelectric nanoparticles, Physical Review B, 90 (2014) 214103.

[60] Anna N. Morozovska and Maya D. Glinchuk, Reentrant phase in nanoferroics induced by the flexoelectric and Vegard effects, J. Appl. Phys., 119 (2016) 094109.

[61] E.A.Eliseev, A.V.Semchenko, Y.M.Fomichov, M. D. Glinchuk, V.V.Sidsky, V.V.Kolos, Yu.M.Pleskachevsky, M.V.Silibin, N.V.Morozovsky, A.N.Morozovska, Surface and finite size effects impact on the phase diagrams, polar and dielectric properties of $(Sr,Bi)Ta_2O_9$ ferroelectric nanoparticles, J. Appl. Phys., 119 (2016) 204104.

[62] Hong, Sahwan, Taekjib Choi, Ji Hoon Jeon, Yunseok Kim, Hosang Lee, Ho-Young Joo, Inrok Hwang et al., Large Resistive Switching in Ferroelectric BiFeO3 Nano-Island Based Switchable Diodes, Advanced Materials, 25 (2013) 2339-2343.

[63] Sakamoto, Takuya, Koichi Okada, Azusa N. Hattori, Teruo Kanki, Alexis S. Borowiak, Brice Gautier, Bertrand Vilquin, and Hidekazu Tanaka, Epitaxial inversion on ferromagnetic (Fe, Zn) $3O_4$/ferroelectric $BiFeO_3$ core-shell nanodot arrays using three dimensional nano-seeding assembly, Journal of Applied Physics, 113 (2013) 104302.

[64] Zhang, Xingang, Bo Wang, Xiuzhang Wang, Xiangheng Xiao, Zhigao Dai, Wei Wu, Junfeng Zheng, Feng Ren, and Changzhong Jiang, Preparation of M@ BiFeO3 Nanocomposites (M= Ag, Au) Bowl Arrays with Enhanced Visible Light Photocatalytic Activity, Journal of the American Ceramic Society, 98(2015) 2255-2263.

[65] A. K. Tagantsev and G. Gerra, Interface-induced phenomena in polarization response of ferroelectric thin films, J. Appl. Phys., 100 (2006) 051607.

[66] A. K. Tagantsev and G. Gerra, Interface-induced phenomena in polarization response of ferroelectric thin films, J. Appl. Phys., 100 (2006) 051607.

[67] M. Fiebig, Revival of the magnetoelectric effect, J Phys D: Appl Phys, 38 (2005) 123-152.

[68] M.D. Glinchuk, E.A. Eliseev, A.N. Morozovska, R . Blinc, Giant magnetoelectric effect induced by intrinsic surface stress in ferroic nanorods, Phys Rev B, 77 (2008) 024106.

[69] D. Rahmedov, S. Prosandeev, J. Íñiguez, L. Bellaiche, Magnetoelectric signature in the magnetic properties of antiferromagnetic multiferroics: Atomistic simulations and phenomenology, Phys Rev B., 88 (2013) 224405.

[70] M.D. Glinchuk, E.A. Eliseev, Y. Gu, L-G Chen, V. Gopalan, A.N. Morozovska, Electric-field induced ferromagnetic phase in paraelectric antiferromagnets, Phys Rev B, 89 (2014) 1014112.

[71] M.D. Glinchuk, E.A. Eliseev, A.N. Morozovska, New room temperature multiferroics on the base of single-phase nanostructured perovskites, J Appl Phys, 116 (2014) 054101.

[72] V. V. Laguta, A. N. Morozovska, E. A. Eliseev, I. P. Raevski, S. I. Raevskaya, E. I. Sitalo, S. A. Prosandeev, and L. Bellaiche, Room-temperature paramagnetoelectric effect in magnetoelectric multiferroics $Pb(Fe_{1/2}Nb_{1/2})O_3$ and its solid solution with $PbTiO_3$, Journal of Materials Science, 51 (2016) 5330-5342.

[73] S. Prosandeev, I.A. Kornev, L. Bellaiche, Magnetoelectricity in $BiFeO_3$ films: First-principles based computations and phenomenology, Phys Rev B, 83 (2011) 020102.